\documentclass[letter]{aa} 

\def\aap{A\&A}

\usepackage{natbib}
\usepackage{graphicx}
\usepackage{fancyhdr}  
\usepackage{url} 

\usepackage{color}

\begin{document} 
   \title{The peculiar fast-rotating star 51 Oph probed by VEGA/CHARA}
%
  \author{N. Jamialahmadi \inst{1,2}
          \and
          P. Berio\inst{2}
          \and 
          A. Meilland\inst{2} 
          \and 
          K. Perraut\inst{3}
          \and 
          D. Mourard\inst{2}
          \and
          B. Lopez\inst{2} 
          \and
          P. Stee\inst{2}
          \and 
          N. Nardetto\inst{2}
          \and 
          B. Pichon \inst{2}
          \and 
          J.M. Clausse \inst{2}
          \and 
          A. Spang \inst{2}
          \and
          H. McAlister\inst{4,5} 
          \and 
          T. ten Brummelaar\inst{5}
          \and 
          J. Sturmann \inst{5}
          \and 
          L. Sturmann \inst{5}
          \and 
          N. Turner \inst{5}
          \and 
          C. Farrington \inst{5}
          \and 
          N. Vargas \inst{5}
          \and 
          N. Scott \inst{5}}           

  \institute{European Southern Observatory, Karl-Schwarzschild-Str. 2, D-85748 Garching, Munich, Germany\\
  \email{njamiala@eso.eu} \hspace{0.1cm} and \hspace{0.1cm} {jami@oca.eu} 
  \and
  Laboratoire Lagrange, UMR 7293, Univ. Nice Sophia-Antipolis, CNRS, Observatoire de la C\^ote d'Azur, 06304 Nice, France   
   \and
    Institut d'Astrophysique et de Plan\'etologie de Grenoble, CNRS-UJF UMR 5571, 414 rue de la Piscine, 38400 St-Martin d'H\`eres, France
    \and
    Georgia State University, PO Box 3969, Atlanta GA 30302-3969, USA
    \and
    CHARA Array, Mount Wilson Observatory, 91023 Mount Wilson CA, USA
    }
  \abstract
    {Stellar rotation is a key in our understanding of both mass-loss and evolution of intermediate and massive stars. It can lead to anisotropic mass-loss in the form of radiative wind or an
excretion disk.}
   {We wished to spatially resolve the photosphere and gaseous environment of 51 Oph, a peculiar star with a very high v\,sin\,i of 267km\,s$^{-1}$ and an evolutionary status that remains unsettled. It has been classified by different authors as a Herbig, a $\beta$ Pic, or a classical Be star.}
   {We used the VEGA visible beam combiner installed on the CHARA array that reaches a submilliarcsecond resolution. Observation were centered on the H$\alpha$ emission line.}
    {We derived, for the first time, the extension and flattening of 51 Oph photosphere. We found a major axis of $\theta_{{\mathrm{eq}}}$=8.08$\pm$\,0.70\,$R_\odot$ and a minor axis of $\theta_{{\mathrm{pol}}}$=5.66$\pm$\,0.23\,$R_\odot$ . This high photosphere distortion shows that the star is rotating close to its critical velocity. Finally, using spectro-interferometric measurements in the H$          \alpha$ line, we constrained the circumstellar environment geometry and kinematics and showed that the emission is produced in a 5.2$\pm$2\,R$_\star$ disk in Keplerian rotation.}
     {From the visible point of view, 51 Oph presents all the features of a classical Be star: near critical-rotation and double-peaked H$\alpha $ line in emission produced in a gaseous disk in Keplerian rotation. However, this does not explain the presence of dust as seen in the mid-infrared and millimeter spectra, and the evolutionary status of 51 Oph remains unsettled. }

   
   \maketitle
%

\section{Introduction}

Many hot stars are observed to be rotating with equatorial velocities higher than 120 km $ s^{-1} $ \citep{1995ApJS...99..135A, 2002ApJ...573..359A}. This fast stellar rotation can have strong effects on the observed stellar shape and the intensity distribution (gravity darkening). Some hot stars show emission lines in their spectra. These features indicate that a gaseous circumstellar environment surrounds these stars. 

The study of the stellar photosphere and circumstellar environment of fast-rotating stars is limited by the lack of spatial resolution of single telescopes. An important and reliable way to extract this information is through long-baseline optical interferometry, which allows us to study the detailed stellar surface properties and the close-by environment through emission lines. Several rapid rotators have been studied using these techniques, including Altair, Vega, Achernar, Alderamin, Regulus, and Rasalhague \citep[respectively]{2001AAS...198.6311V,2006ApJ...645..664A,2006Natur.440..896P, 2003A&A...407L..47D,2007Sci...317..342M}.

This paper is dedicated to the study of 51 Oph: a rapidly rotating star, $v\hspace{0.1cm} \sin \hspace{0.1cm} i$=267$ \pm 5 $ km.s$^{-1} $\citep{1997MNRAS.286..604D} with an age of 0.3 Myr and a mass of $ \sim $4 {$M_{\odot}$} (Van den Ancker et al. 1998). This star appears to be a peculiar source in an unusual transitional state. Compared with Herbig Ae/Be stars, \citet{2001A&A...369L..17V} and
\citet{2001A&A...365..476M} confirmed by analyzing ISO spectra that 51 Oph has more gas than other HAeBe stars. In addition, Malfait et al. (1998) and Leinert et al. (2004) found that the dusty disk of 51 Oph is much smaller ($<$2AU) than the disk of most HAeBe stars.

Compared with $ \beta $~Pic-like stars, 51 Oph does not exhibit a far-infrared-excess bump, which is inconsistent with the presence of an outer dusty disk \citep{1998A&A...331..211M}. \citet{2013A&A...557A.111T} suggested a difference between 51 Oph and $ \beta $~Pic because the 51 Oph dust grains can be primordial, while the dust grains in an exozodiacal ring around $ \beta $~Pic have to be replenished. Furthermore, 51 Oph does exhibit an $ H_{\alpha} $ line in emission, which is not the case of $ \beta $~Pic stars. \citet{2008PhDT.........3B} and  \citet{2008A&A...489.1151T} concluded that 51 Oph is most likely a classical Be star surrounded by a compact gaseous disk. This scenario agrees well with the following properties reported in previous studies: high rotational velocity, most of the near-infrared continuum emission arising from the gaseous disk, and hot CO gas in a rotating Keplerian disk. However, the remaining presence of dust around 51 Oph is not supported by this scenario and suggests that this star follows an intriguing evolution scheme.

We here report new elements that lend credit to the hypothesis that 51 Oph is a Be star. Hereafter, we present observations of 51 Oph with a spectro-interferometer operating at optical wavelengths, the VEGA instrument \citep{2009A&A...508.1073M,2011A&A...531A.110M} installed at the CHARA Array \citep{2005ApJ...628..453T}. This unique combination of spectrally and spatially resolved information allowed us to resolve the photosphere of the star and to probe the kinematics of the gaseous component producing the H$ \alpha $ line on sub-AU scales.

\begin{table}
\caption{CHARA/VEGA observing log of 51 Oph. For the continuum observations, the calibrators HD 150366 with the uniform disk angular diameter $ \theta_{{\mathrm{UD}}}$=0.28 $\pm$ 0.02 mas and $ \theta_{{\mathrm{UD}}}$=0.39 $\pm$ 0.03 mas for HD 163955 were used. For the observations in the $ H_{\alpha} $ line, HD 170296 with $ \theta_{{\mathrm{UD}}}$=0.43 $\pm$ 0.03 mas was used. The predicted uniform disk angular diameter (in mas) in R band were derived from the JMMC SearchCal software \citep{2006A&A...456..789B}.}

\label{table}
{ \renewcommand{\arraystretch}{1.} 
\setlength{\tabcolsep}{4pt}   
\centering
\begin{tabular}{ |c   c|  |c  c | }
\hline
\multicolumn{2}{|c}{ {\bf{Continuum}}} & \multicolumn{2}{c|}{\bf{$ H_{\alpha} $ line}} \\ 
\hline
Obs. time & Telescopes & Obs. time& Telescopes\\ [0.5ex] 
(UTC) &conf. &(UTC)& (conf. ) \\ [0.5ex]
\hline\hline
 2014-07-04 05:40&  E2-S2-W2&--------- &-----\\
 2014-07-04 06:15&  E2-S2-W2&2013-05-25 07:12&E1-E2\\
 2014-07-04 06:49&  E2-S2-W2&2013-07-28 03:39&S1-S2 (1)\\
 2014-07-08 05:53&  E2-S2-W2&2013-07-28 05:43&S1-S2 (2)\\
 2014-07-08 06:18&  E2-S2-W2&2014-05-02 10:04&W1-W2 (1)\\
 2014-07-08 06:43&  E2-S2-W2&2014-05-02 10:47&W1-W2 (2)\\
\hline

\end{tabular}}
\end{table}

   \section{Observations and data processes}
  
   Observations were performed around the H$ \alpha $ line, between 630 and 670 nm, using the VEGA  medium spectral resolution (MR)(R=5000) to study the gaseous disk by combining three pairs of telescopes: W1W2, E1E2, and S1S2. The observation log is given in Table 1.
     The source 51 Oph was also observed in the continuum at 700 nm using the triplet E2S2W2 to study its photosphere. The (u,v) plane coverage corresponding to these data is presented in the left panel of Fig. 1 in continuum and in the right panel in the H$ \alpha $ line.

\begin{figure}[htb]
  \begin{center}
  \setlength{\unitlength}{1cm}
  \begin{picture}(12, 4)
     \put(0,0) {\includegraphics[width=5cm, height=4cm]{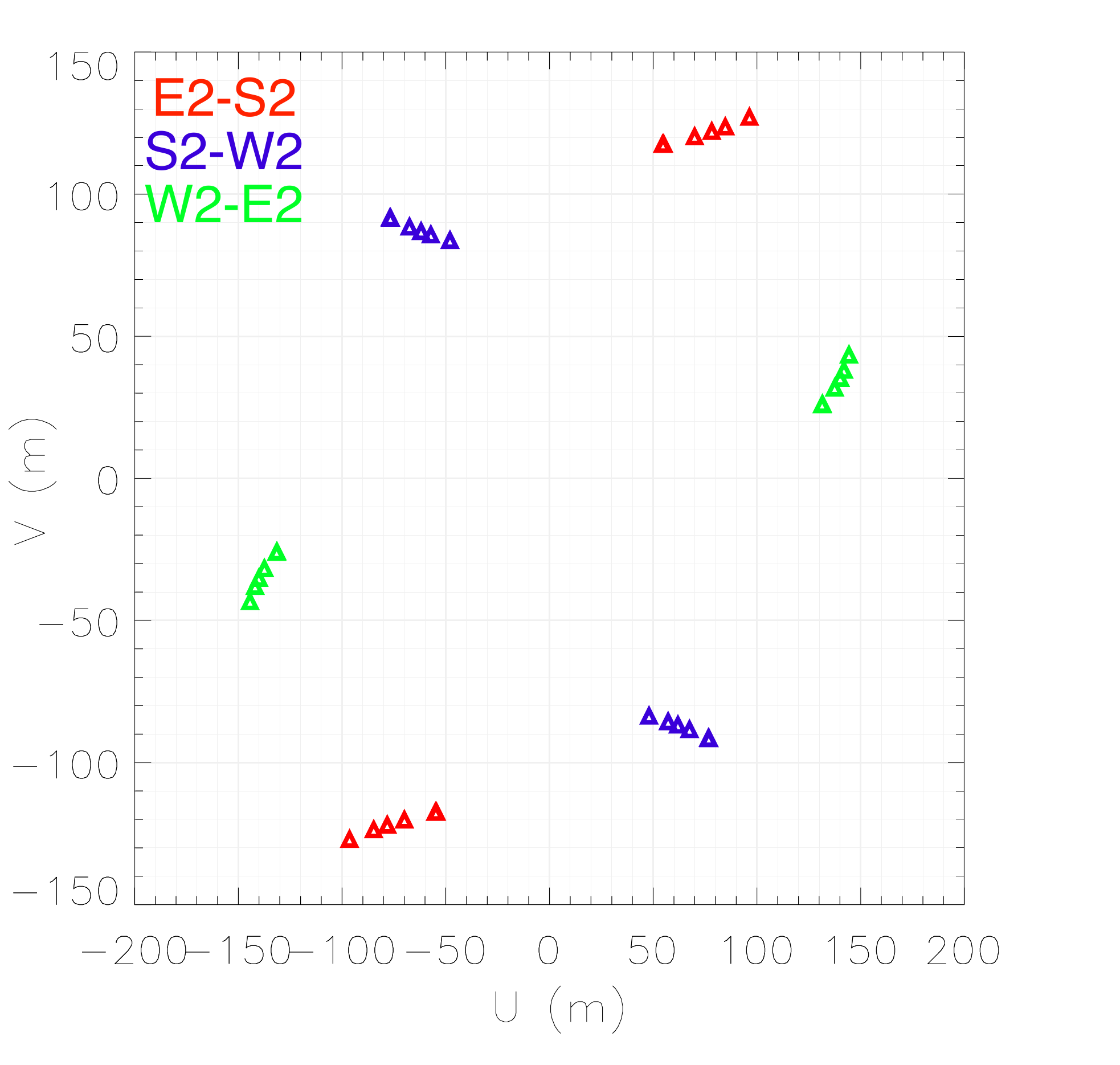}}
     \put(4.5,0) {\includegraphics[width=4.6cm, height=3.8cm]{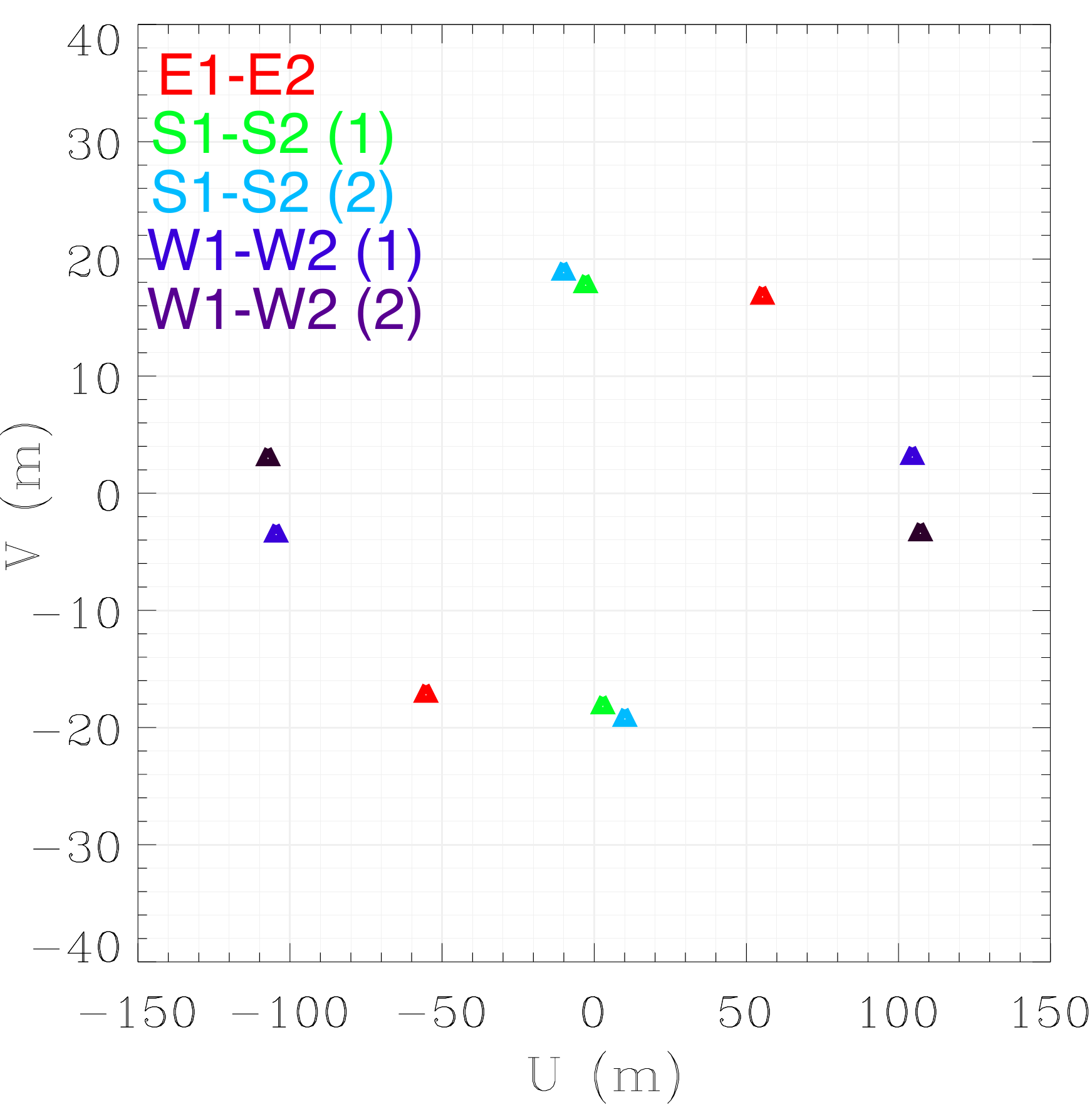}}
      \end{picture}
   \caption{ (u,v) coverage obtained for the observations of 51 Oph in the continuum (left) and the in the H$ \alpha $ line (right).}
  \end{center}
  \label{Lable}
 \end{figure}

The data processing of VEGA is composed of two parts. The data in the continuum are processed using the power spectral method giving the squared visibilities \citep [see][for details]{2009A&A...508.1073M}. For this analysis, we used a spectral bandwidth of 15 nm. The processing of the data in the spectral line is based on the cross-spectrum method, which provides differential visibilities and phases across the line. This method was applied with a spectral bandwidth of 0.2 nm.

\begin{figure}[h]
        \centering
        \includegraphics[width=7cm, height=7cm]{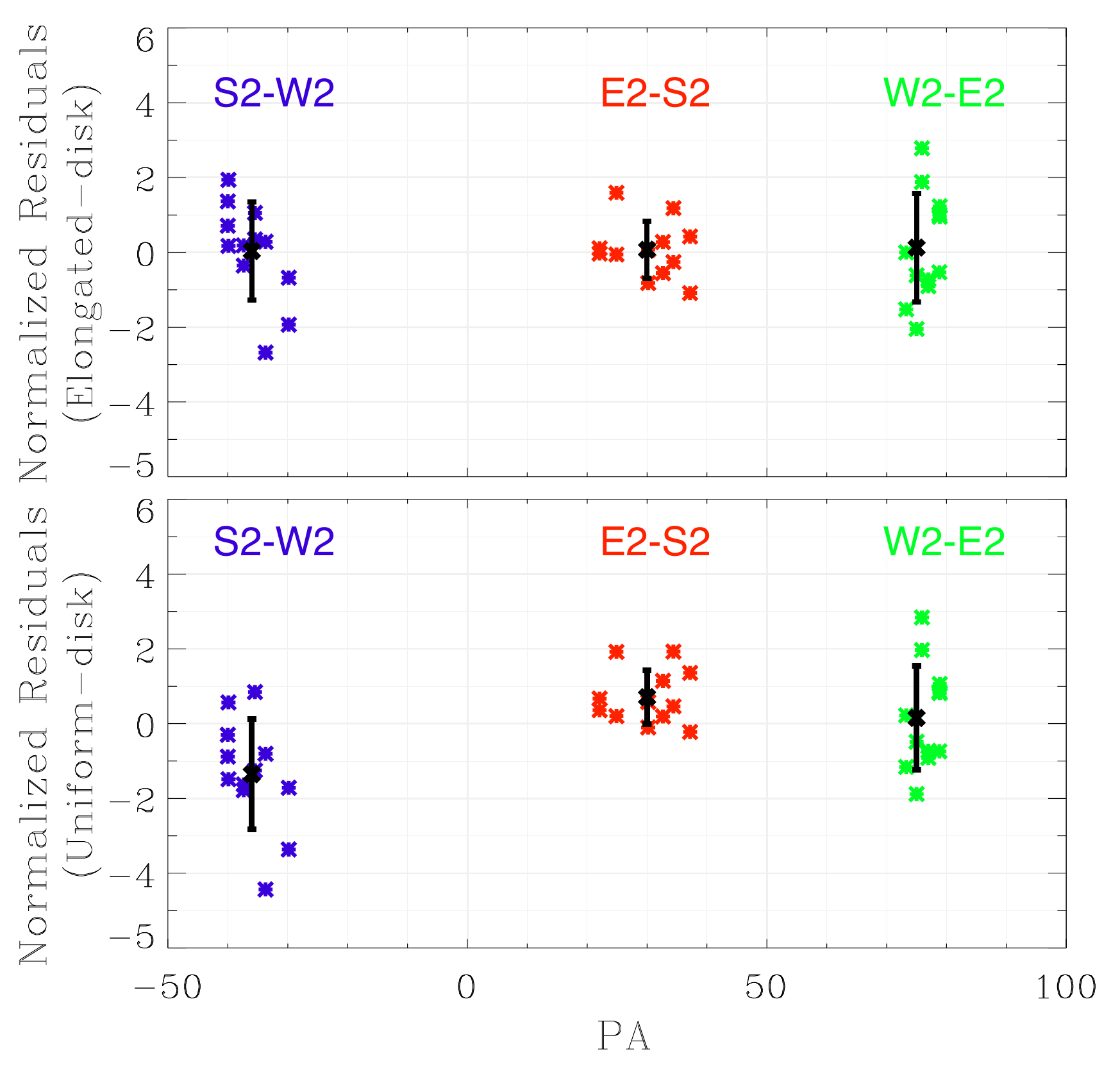}
        \caption{The normalized residuals between the continuum data and an elongated disk model (top) and a uniform disk model (bottom) are represented for three pairs of telescopes. The black points represent the residual average for each baseline. }
        \label{Fig:2}%
\end{figure} 
\section{Continuum emission: stellar photosphere}
 We used the LITpro2 model-fitting software for optical-to-infrared interferometric data developed by the Jean-Marie Mariotti Center (JMMC) to interpret our data. Figure 2 shows that the elongated disk model is more consistent with the data than the uniform disk model for the stellar photosphere for the baselines S2-W2 and E2-S2: the mean of the residuals of S2-W2 and E2-S2 is not equal to zero for the uniform disk model. However, both models have nearly the same normalized residuals for the baseline W2-E2. According to Table 2, we found the major axis, $ \theta_{\mathrm{eq}} $, the minor axis, $ \theta_{\mathrm{pol}} $, the elongated ratio, $ \theta_{\mathrm{eq}} $/$ \theta_{\mathrm{pol}} $ , and the major axis orientation, PA, using an elliptical model. Using Hipparcos 2007 data \citep{2007A&A...474..653V} and calculating the distance from the parallax (d=124 $ \pm 4 $ pc), we derived an equatorial radius $ R_{\mathrm{eq}} $ = 8.08 $ \pm $ 0.7 $R _{\odot} $ and a polar radius $ R_{\mathrm{pol}} $ = 5.66 $ \pm $ 0.23 $R _{\odot} $ from $ \theta_{\mathrm{eq}} $ and $ \theta_{\mathrm{pol}} $ , respectively. We assumed the continuum flux to exclusively
come from the central star. 
A comparison of the H$ \alpha $ emission line of 51 Oph with classical Be stars shows that the intensity level of 51 Oph is low (only 1.2 time above the continuum, see Fig. 3). The flux ratio between the disk and the central star is therefore lower than in classical Be stars. In classical Be stars, this flux ratio in the continuum is between 5 $\%$ and 20 $\%$ in the visible. We therefore assumed that the disk contribution in the continuum is negligible. The fit of an elliptical model for the stellar photosphere produces a reduced $ \chi^{2} $ ($ \chi_{r}^{2} $) of 1.52.

\begin{table}
\caption{ Best-fit parameters obtained from the elliptical model for the stellar photosphere. $ \chi_{r}^{2} $ is the total reduced $ \chi^{2} $}
\centering          
\begin{tabular}{ c  c }     
\hline      
    Parameter&Value\\ [0.5ex]
\hline\hline                
   Major axis of $ \theta_{\mathrm{eq}} $&0.6$ \pm $0.05 mas\\[0.5ex]
   Minor axis of $ \theta_{\mathrm{pol}} $&0.42$ \pm $0.01 mas\\[0.5ex]
   Elongated ratio ($ \theta_{\mathrm{eq}} $/$ \theta_{\mathrm{pol}} $) &1.45  $ \pm $ 0.12\\[0.5ex]
   Position angle (PA)&$138 $ $ \pm $  $3.9^ \circ $\\ [0.5ex]
\hline\hline
$ \chi_{r}^{2} $&  1.52 \\ [0.5ex]
\hline
 \end{tabular}
\end{table}

 \section{H$ \alpha $ emission line}
 \subsection{Qualitative analysis}
The differential visibility and phase for the five baselines are plotted in Fig. 3. The visibility exhibits a drop in the emission line that is caused by a variation of the circumstellar environment extension and relative flux between the continuum and the line, as already explained in \citet{2007A&A...464...59M}. The phase shows S-shaped variations in the line. This is characteristic of a rotating equatorial disk, as described in \citet{2012A&A...538A.110M}. For the longest baseline, W1W2 the disk is probably over-resolved, so that the differential phase deviates from the S shape, as explained in \citep{2012A&A...538A.110M}. The VEGA spectrum of 51 Oph obtained during our observations exhibits an H$ \alpha $ emission line with a double-peak similar to the spectrum found by \citet{1997MNRAS.290..165D}, which reflects the presence of the circumstellar disk. H$ \alpha $ profile wings clearly show that the disk is seen nearly edge-on.

\subsection{Kinematic model}

To model the wavelength dependence of the visibility, the differential phase and the intensity profile of H$ \alpha $ line emission, we used a simple kinematic model of an expanding and/or rotating thin equatorial disk\footnote{As input of our model, we used a synthetic absorption in $H_{\alpha}$ line calculated with the TLUSTY code \citep{1998ASPC..138..139H}.}. This model has prviously been used for some classical Be stars, for instance, by \citet{2011A&A...532A..80M} and \citet{2012A&A...538A.110M} and is described in detail by \citet{2011A&A...529A..87D}. Assuming no expansion, the model free-parameters can be classified into three categories:

\begin{enumerate}
\item[1.] the geometric parameters: stellar radius ($R_{\mathrm{\star}} $), distance (d), inclination angle (i), and disk major-axis position angle (P.A.),
\item[2.] the kinematic parameters: rotational velocity ($V_{\mathrm{rot}}$) at the disk inner radius (i.e., photosphere) and the rotation ($ \beta $) velocity laws,
and\item[3.] the disk emission line parameters: disk FWHM in the line (a) and the line equivalent width (EW).
\end{enumerate}


\begin{table}
\caption{Best-fit parameters obtained from our simple kinematic model of the circumstellar disk.}
{ \renewcommand{\arraystretch}{1.2} 
\setlength{\tabcolsep}{1.5pt}   
\centering          
\begin{tabular}{ c  c  c }     
\hline      
    Parameter&Value&Remarks\\ [0.5ex]
\hline\hline  
& \bf{Geometric parameters}  &   \\                  
   $ R_{\mathrm{eq}} $\tablefootmark{\ast}&8.08$R _{\odot} $& This work (Sect 4)\\[0.5ex]
   d\tablefootmark{\ast}&124 pc& (van Leeuwen 2007)\\[0.5ex]
   i&${80^{+7} _{-30}} $ deg& This work\\ [0.5ex]
   PA&$ {122^{+15} _{-25}} $ deg& This work\\ [0.5ex]
\hline
& \bf{Kinematic parameter}  &   \\ 
$ V_{\mathrm{rot}} $\tablefootmark{\ast}&267 km.s $ ^{-1} $&  (Dunkin et al. 1997a) \\ [0.5ex]
\hline
& \bf{$ H_{\alpha} $ line parameters} &   \\ 
a&${5.6\pm 2}$ $ R_{\star} $& This work\\ [0.5ex]
EW&${5.0^{+2} _{-3}}$ ${\AA} $& This work\\ [0.5ex]
\hline\hline
$ \chi_{r}^{2} $&  0.7 & \\ [0.5ex]
\hline
 \end{tabular}}
 \tablefoottext{\ast}{Fixed parameters.}
\end{table}

The star distance was calculated following \citet{2007A&A...474..653V}. $R_{\mathrm{eq}} $, the equator stellar radius, was derived from the fit of the elliptical model according to Sect. 3 (in solar radius). We assumed that the disk is directly connected to the stellar surface, therefore the rotational velocity ($V_{\mathrm{rot}}$) should be equal to the stellar rotational velocity. We also assumed that the gaseous disk surrounding 51 Oph is Keplerian, so that we set $ \beta $, the rotation velocity law, to -0.5. The other four parameters were left free.

\begin{figure*}[htb]
  \begin{center}
  \setlength{\unitlength}{1cm}
  \begin{picture}(20, 7.4)
     \put(0,0) {\includegraphics[width=12.7cm, height=7.4cm]{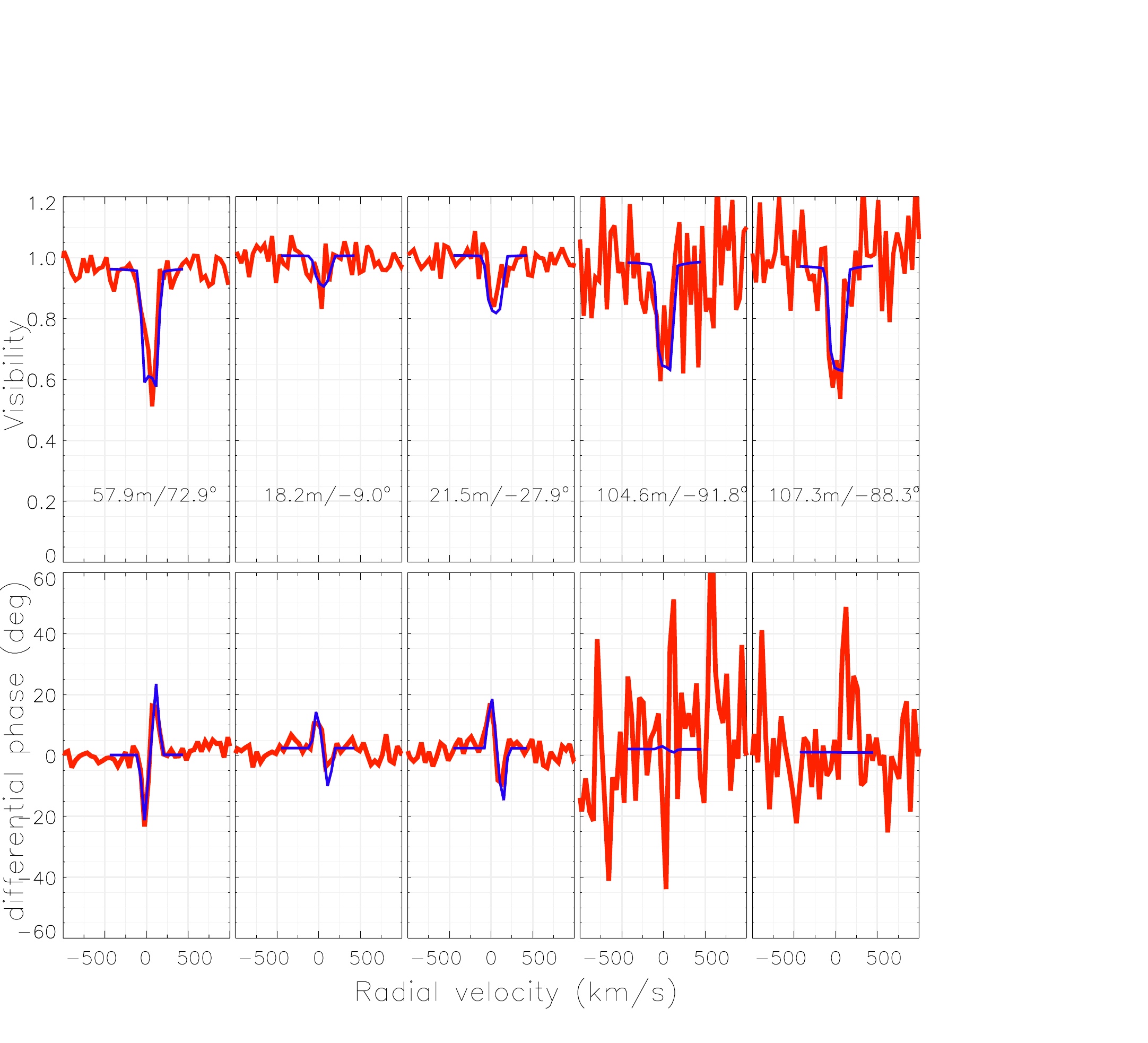}}
     \put(12.8,2.85) {\includegraphics[width=4.5cm, height=4.5cm]{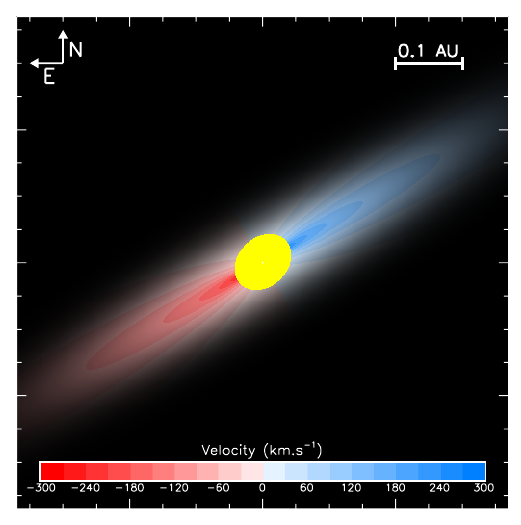}}
     \put(12.4,0.3) {\includegraphics[width=5.55cm, height=2.6cm]{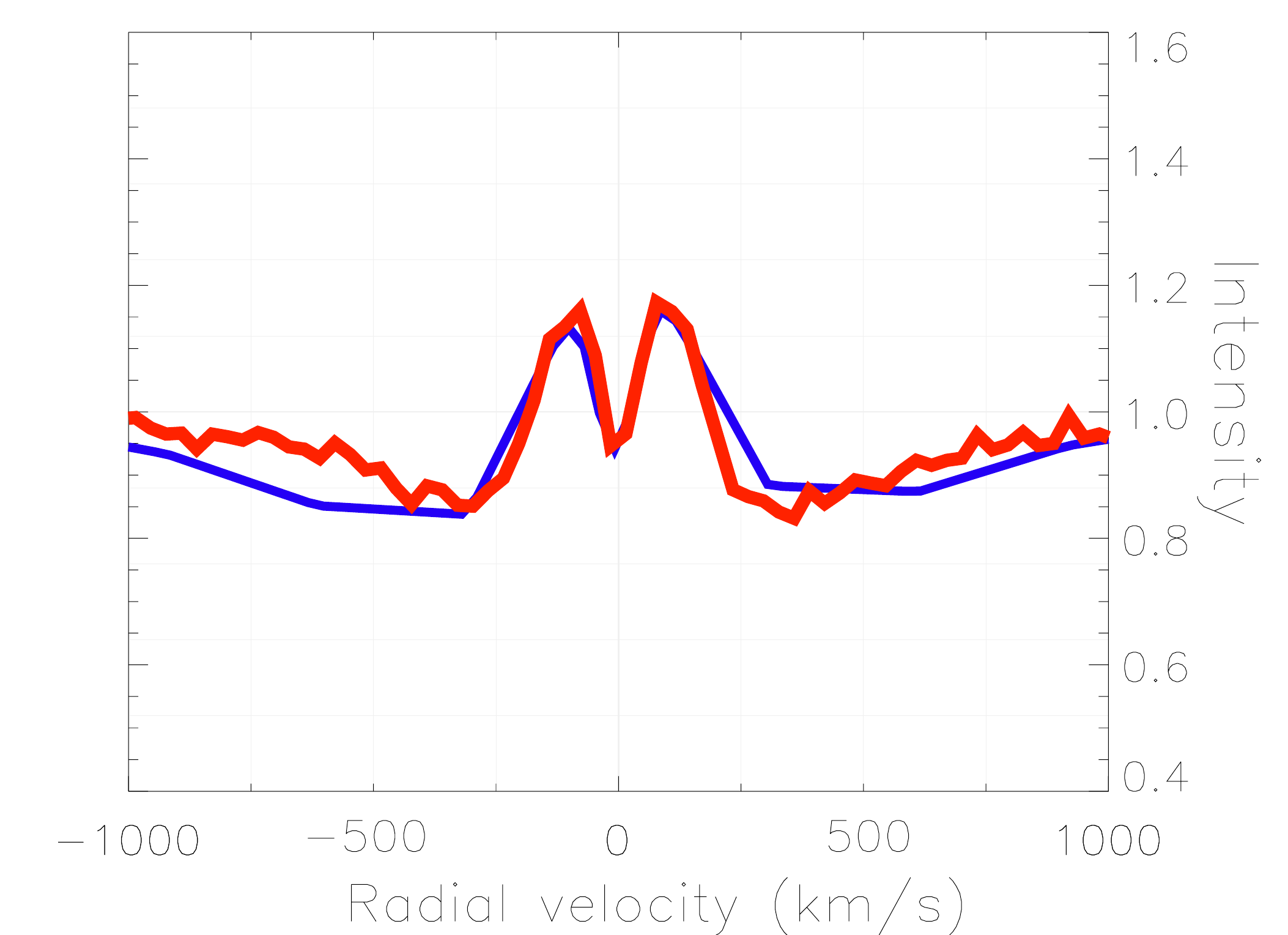}}
      \end{picture}
   \caption{Visibility and differential phase variations in the H$ \alpha $ emission line and in the surrounding continuum obtained from our medium spectral-resolution (MR) CHARA/VEGA observations (red). The best-fit kinematic model for this dataset is overplotted in blue. The plot (right bottom panel) corresponds to the H$ \alpha $ intensity profile. The top right panel of the figure shows the elliptical model for the central star 51 Oph and our kinematic model for a purely rotating disk.}
  \end{center}
  \label{Lable}
 \end{figure*}

For our target we computed several hundreds of models to constrain the free parameters, determined the uncertainties, and tried to detect any degeneracy or linked parameters.
To reduce the number of computed models, we started with a qualitative estimation of the parameters from our interferometric data (P.A., i, a, and EW) and explored the parameter space with decreasing steps to converge to the $ \chi^{2} $ minimum ($ \chi_{r}^{2} $). The final parameter values for the best-fit models are presented in Table 3. The corresponding differential visibilities and phases are overplotted in Fig 3. The final reduced $\chi_{r}^{2}$ is 0.7.

The inclination angle found for the gaseous disk in our study is compatible with the previous estimate of \citet{2005A&A...430L..61T}, that is, ${{88^{\circ}}^{+2} _{-35}}$ and that of \citet{2008A&A...489.1151T}, ${{82^{\circ}}^{+8} _{-15}}$. Moreover, the disk position angle that we derived, that is, $ {{122^{\circ}}^{+15} _{-25}}$ , is also compatible with that of \citet{2008A&A...489.1151T}, $ {{126^{\circ}}^{+15} _{-5}}$ , and is likewise consistent with the angle found by \citet{2009ApJ...703.1188S}, $ {{126^{\circ}}^{+0.5} _{-0.15}}$. Finally, the position angle of the gaseous disk in our study also agrees well with the position angle found for the major
axis of the central star, that is, $138 \pm $  $3.9^{\circ}$, showing that the disk is in the equatorial plane.




\section{Discussions} 

The first main result from our  CHARA/VEGA observing campaign on 51 Oph is the measurement of the highly flattened stellar photosphere. Assuming that the photosphere distortion is due to the high rotational velocity of the star, the stellar rotational rate can be inferred. Assuming a Roche model and for an edge-on star, a 1.45$\pm$0.12 distortion corresponds to an angular rotation rate of $\Omega$/$\Omega_c$=0.99$\pm$0.02, or a linear one of V/V$_c$=0.96$\pm$0.1. Another method for deriving the rotational rate is to use the mass and radius of the star and the measured v\,sin\,i. The critical velocity is given by $V_c=\sqrt{GM_{\star}/R_{e}}=\sqrt{2GM_{\star}/3R_{p}} $ , where the 'e' and 'p' subscripts correspond to the equatorial and polar values. Knowing the mass of 51 Oph, 4$M _{\odot} $ (van den Ancker et al. 1998) and $ R_{\mathrm{eq}} $, 8.08$ R_{\odot} $, the calculated critical velocity is 308 km.s $ ^{-1} $. Neglecting the stellar inclination angle, the rotational velocity of the star, that is, 267 km.s $ ^{-1} $ , is 86 $ \% $ of the critical one in linear velocity. As the star might not be seen fully edge-on, this represents a lower limit of the rotational rate. These two different methods give coherent results, and it is clear that 51 Oph is very close to critical rotation. 

Assuming critical rotation for a star with a rotational velocity of 267\,km\,s$^{-1}$ and a polar radius of R$_p$=5.66$\pm$0.23\,R$_\odot$, we can deduce a mass of 3.3$\pm$0.1\,M$_\odot$. This is 25$\%$ lower than the estimate of van den Ancker et al. (1998), who assumed pre-main sequence evolutionary tracks, that is, 4\,M$_\odot$, and 15$\%$ lower than the estimate of van den Ancker (2001),
who  assumed post-main sequence evolutionary tracks, that is, 3.8\,M$_\odot$. Nevertheless, even if our measurement favors the post-main sequence hypothesis, the difference between the two estimated masses is too small to conclude on the evolutionary
status of 51 Oph evolutionary.

The second important result of our study is the direct measurement of the polar radius. Considering the estimate of the effective temperature by Van den Ancker et al. (1998) of 10000\,K, it is too high for the star to be classified as a main-sequence star. The standard value for a B9.5V to A0V star is about 2.4-2.6\,R$_\odot$. Accordingly, 51 Oph should be classified as a giant star. This distance from the main sequence, which was already found by Ancker et al. (1998), was used to compute the age, assuming that 51 Oph was a pre-main sequence star still in the contraction phase. Nevertheless, as we stated above, this value is also compatible with a post-main sequence star. 

The last main result from our CHARA/VEGA observing campaign concerns the circumstellar environment geometry and kinematics. We first note that the extension in the H$\alpha$ line, 0.20$\pm0.07$AU, is compatible with the one found by \citet{2008A&A...489.1151T} in the K-band CO lines using the VLTI/AMBER instrument, 0.15$^{+0.07}_{-0.04}$ AU. This emission is also lower than the dust sublimation radius of 51 Oph computed using Eq. 1 from Dullemond and Monnier (2010), 1.17 AU (i.e., with T$_{eff}$=10000\,K T$_{rim}$=1500\,K and R$_\star$=R$_p$). With our kinematic model, we showed that the spectro-interferometric data in the H$\alpha$ emission line are fully compatible with a geometrically thin disk in Keplerian rotation. We note that the differential phase variations cannot be reproduced with an expanding wind. The disk features are similar to those of the few classical Be stars studied with the same instrument (\citealt{2011A&A...532A..80M}, \citealt{2011A&A...529A..87D}), even if the extension in H$\alpha$ is slightly smaller in the case of 51 Oph. On the other hand, the emission could come from an accretion disk whose geometry and kinematics would not differ strongly from the one of excretion disk of classical Be stars. The main difference between the two models would be the radial velocity, positive for the excretion disk and negative for the accretion disk. However, this low radial velocity of about 0.1 to 1\,km\,s$^{-1}$ is negligible compared to the high rotational velocity in the disk and cannot be constrained in our data.

Finally, from the visible point of view, 51 Oph presents all the features of a classical Be star: critical rotation, double-peaked H$\alpha $ line in emission, and a circumstellar gas disk of a few stellar radii in Keplerian rotation. However, this does not explain the presence of dust as seen in the mid-infrared and millimeter range, and the evolutionary status of 51 Oph still remains unclear. Nevertheless, even if 51 Oph were a pre-main-sequence star, it would make a perfect progenitor for a classical Be star
because it is very unlikely that it will lose enough angular momentum to significantly slow down during its on-going contraction phase.

\begin{acknowledgements}
 The CHARA Array is operated with support from the National Science Foundation through grant AST- 1211129, the W. M. Keck Foundation, the NASA Exoplanet Science Institute, and from Georgia State University. This research has made use of the SearchCal service of the Jean-Marie Mariotti Center and of CDS Astronomical Databases SIMBAD and VIZIER. N.J. warmly thanks all the VEGA observers that permitted the acquisition of this set of data. N.J. also acknowledges the Ph.D. financial support from the Observatoire de la C\^ote d'Azur and the PACA Region. We thank T. Lanz for useful discussions. This research has largely benefited from the advices of our colleague Olivier Chesneau, who passed away this Spring. The authors wish to acknowledge his help.
     
\end{acknowledgements}

\bibliographystyle{aa}
\bibliography{Letter_new}

\end{document}